\documentclass[letterpaper]{article}
\usepackage{aaai23}  
\usepackage{times}  
\usepackage{helvet}  
\usepackage{courier}  
\usepackage[hyphens]{url}  
\usepackage{graphicx} 
\usepackage{natbib}  
\usepackage{caption} 
\title{Robust-MSA: Understanding the Impact of Modality Noise \\on Multimodal Sentiment Analysis}
\author{
    Huisheng Mao, \textsuperscript{\rm 1,\rm 2}
    Baozheng Zhang, \textsuperscript{\rm 1,\rm 3}
    Hua Xu, \textsuperscript{\rm 1,\rm 2}\thanks{\quad Hua Xu is the corresponding author.}
    Ziqi Yuan, \textsuperscript{\rm 1,\rm 2}
    Yihe Liu\textsuperscript{\rm 1,\rm 3}
}
\affiliations{
    \textsuperscript{\rm 1} State Key Laboratory of Intelligent Technology and Systems,
Department of Computer Science and Technology,\\Tsinghua University, Beijing 100084, China\\
    \textsuperscript{\rm 2} Beijing National Research Center for Information Science and Technology(BNRist), Beijing 100084, China\\
    \textsuperscript{\rm 3} School of Information Science and Engineering, Hebei University of Science and Technology, Shijiazhuang 050018, China\\
    xuhua@mail.tsinghua.edu.cn
}

\begin{document}

\maketitle

\begin{abstract}
Improving model robustness against potential modality noise, as an essential step for adapting multimodal models to real-world applications, has received increasing attention among researchers. For Multimodal Sentiment Analysis (MSA), there is also a debate on whether multimodal models are more effective against noisy features than unimodal ones. Stressing on intuitive illustration and in-depth analysis of these concerns, we present Robust-MSA, an interactive platform that visualizes the impact of modality noise as well as simple defence methods to help researchers know better about how their models perform with imperfect real-world data. 

\end{abstract}

\section{Introduction}

Multimodal Sentiment Analysis (MSA) is an increasingly popular task in multimodal machine learning \cite{baltruvsaitis2018multimodal, soleymani2017survey}. It aims to analyze speaker’s sentiment from a short video clip containing three modalities: visual, audio and text. Although researchers have achieved promising improvements over the years \cite{BERT_MAG, MISA}, models of this task are not as widely used in applications as those of other popular machine learning tasks. Lacking of ability to give correct predictions on real-world samples is a major cause. Videos in popular MSA datasets, such as MOSI \cite{MOSI}, MOSEI \cite{MOSEI} and CH-SIMS \cite{CH-SIMS, CH-SIMSv2}, are usually handpicked samples: speakers’ faces are frontal without occlusion; their voices are clear without noise or interruption; the text transcripts are manually revised thus have minimal error. In real-world scenarios, however, such “perfect” samples are not the common case. Speakers may turn away from the camera; their voices may be overwhelmed by environment noise; text transcript, the dominant modality, has to be obtained via Automatic Speech Recognition (ASR) and thus may have devastating errors.

To address these problems, researchers have identified a key challenge in MSA: how to effectively improve model robustness against modality noise \cite{NewSurvey, TPFN}. In order to develop a robust model, it's essential to understand how modality noise affect existing models. In this paper, we present Robust-MSA, an interactive visualization platform for understanding what kind of influence modality noise impose on MSA models.

\section{Demonstrating Robust-MSA}

Robust-MSA takes user-generated videos as input. Speech recognition is proceeded automatically after uploading the video. Manually revising of the generated transcript is needed for obtaining a ``perfect" instance. Robust-MSA then aligns video with the transcript, and offers customization of noise on word granularity. The platform visualizes video-text alignment results for both original and noise-injected version of the video, and highlights how they differ and lead to wrong predictions. Furthermore, Robust-MSA provides visualization of modality features in a timeline view. This helps researchers better understand how noise affect the feature extraction process and lead to mispredictions.

\subsection{Noise Generation}

Modality noise in MSA usually results in several common problems at feature level. For example, occlusion and bad camera angle may cause facial detection failure, which leads to zero values in corresponding feature dimension; noisy environment and bad microphone reception may result in ineffective audio features; ASR algorithm and typos may introduce transcript errors and further lead to incorrect text features. Robust-MSA provides six different noise simulator imitating real-world data imperfections on the word granularity. For video modality, "Blank-Screen" and "Gaussian-Blur" are supported. For audio modality, the platform provides "Mute" method and six different kind of  "additive background noise" from DEMAND dataset \cite{DEMAND}. For text modality, the options are "Replace" and "Remove". These six methods can well simulate most modality noises from real-world scenarios since they result in the same problems at feature level. 

\begin{figure*}[t]
  \centering
  \includegraphics[width=\linewidth]{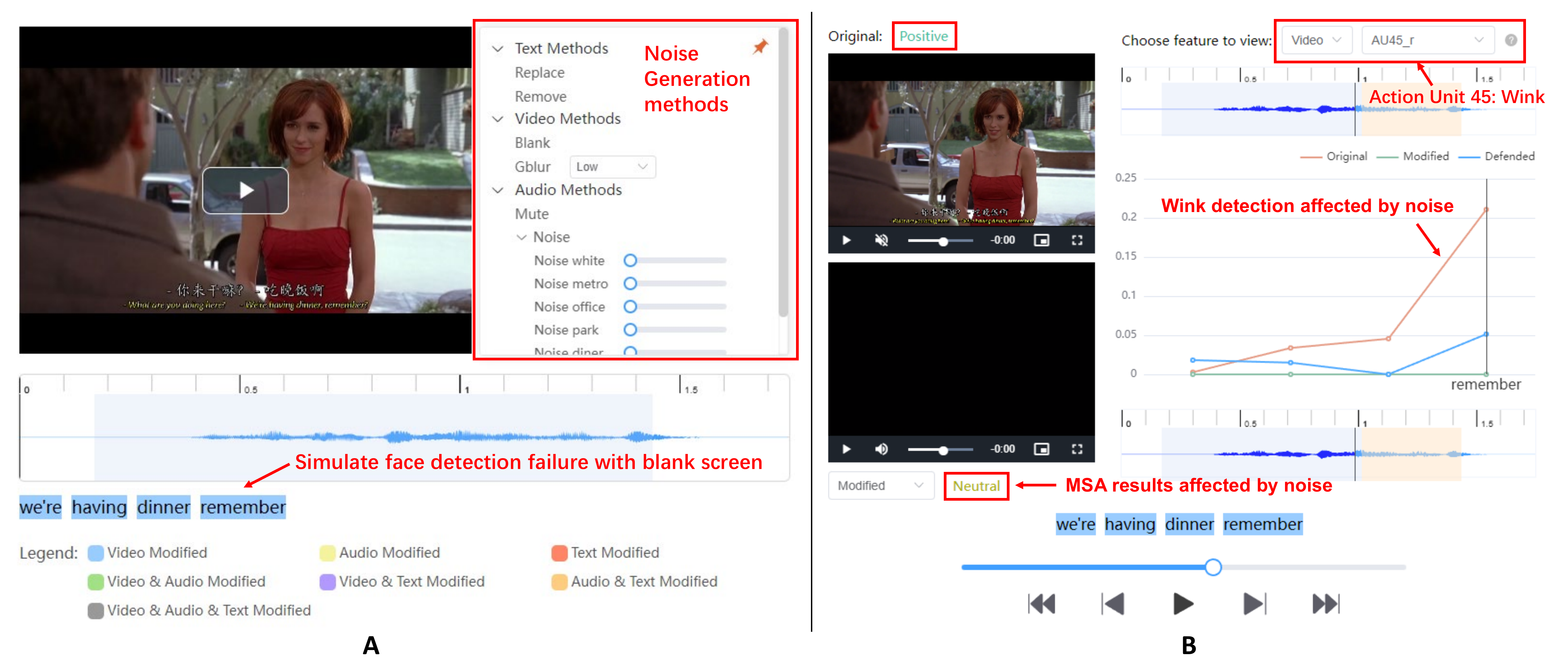}
  \caption{Noise Influence Demonstration. Left: Noise Injection, Right: feature and prediction comparison.}
  \label{fig: intermediate_result_test}
\end{figure*}

As shown in Figure \ref{fig: intermediate_result_test}A, to add modality noise to a video, simply drag one of the six methods, drop it onto a word, and the method will be automatically applied to the corresponding modality of the word. The added noise will be highlighted with different background colors according to their modality. 

\subsection{Noise Defence Methods}

Robust-MSA provides three simple noise defence methods, including audio denoising, video motion compensated interpolation (MCI) at raw data level, and feature interpolation at feature level. The audio denoising method denoises the raw audio wave with Fast Fourier Transform (FFT); the video MCI method generates missing frames with Enhanced predictive zonal search algorithm (EPZS) algorithm \cite{EPZS}; the feature interpolation simply does a linear interpolation on missing features. The defended video and features can be viewed on final result page. Researchers can experiment for themselves to figure out whether these simple defence methods help to improve model robustness or not.

\subsection{End-to-End MSA Pipeline}

\noindent\textbf{Feature Extraction.} For real-time application, identical modality features for both training and inference stage are required. Specifically, eGeMapsv02 \cite{egemaps} feature set is adopted as acoustic features, facial landmarks \cite{landmarks} and Action Units (AU) \cite{action_unit} are extracted as visual features, BERT \cite{BERT} language model is selected to process textual features. Moreover, a pretrained Wav2vec2 model \cite{wav2vec2} is used to generate timestamps for video/audio to text alignment. All above customized feature extraction are performed with the help of MMSA-FET toolkit \cite{M-SENA}.

\noindent\textbf{Integrated MSA Models.} Currently, Robust-MSA supports eight MSA benchmark models including TFN \cite{TFN}, LMF \cite{LMF}, MISA \cite{MISA}, MAG-BERT \cite{BERT_MAG}, Self-MM \cite{self-mm}, MMIM \cite{MMIM} and TFR-Net \cite{TFR-NET} for performances comparison on the noisy environment. All models are trained on MOSEI \cite{MOSEI}. The final sentiment prediction shown in Figure \ref{fig: intermediate_result_test}B, Robust-MSA averages the models' outputs and map the score into three classes, ``Negative", ``Neutral" and ``Positive". 

\subsection{Noise Influence Demonstration}

To help researchers better understand the affect of modality noise on both extracted features and sentiment predictions, Robust-MSA presents the original video, its noised-injected version, and noise-defended version in alignment with the transcript. To show the alignment results, corresponding audio segment and text are highlighted accompanied by video. Users can also click on a word or use the control buttons below to quickly navigate through words in video and audio. With these convenient operations, users can easily pinpoint video and audio segments where simulated modality noise is introduced. 
Moreover, the platform visualizes modality features of three versions of the video in a line chart where the x-axis represents corresponding words in the transcript, as shown in the right of Figure \ref{fig: intermediate_result_test}.

\section{Engaging the Audience}

Our demonstration focus on how even a tiny inconspicuous modality noise, such as facial occlusion in a few frames, may lead to incorrect predictions even for models designed to overcome such noises in MSA tasks. As shown in Figure \ref{fig: intermediate_result_test}A, we modified the original video by dropping the entire visual modality to simulate face detection failure. The results are shown in Figure \ref{fig: intermediate_result_test}B, the noise-injected video is classified as "Neutral" while the original one is "Positive". From feature view we can inspect the ``wink" action unit, which is a crucial visual cue for sentiment prediction.

Hopefully, the demonstration will raise more concerns on this topic and help the audience realize the importance of model robustness to applying MSA models in real-world applications. 

\appendix

\section{Acknowledgments}

This paper is founded by National Natural Science Foundation of China (Grant No. 62173195) and Beijing Academy of Artificial Intelligence(BAAI).


\begin{thebibliography}{23}
\providecommand{\natexlab}[1]{#1}

\bibitem[{Baevski et~al.(2020)Baevski, Zhou, Mohamed, and Auli}]{wav2vec2}
Baevski, A.; Zhou, Y.; Mohamed, A.; and Auli, M. 2020.
\newblock wav2vec 2.0: A framework for self-supervised learning of speech
  representations.
\newblock \emph{Advances in Neural Information Processing Systems}, 33:
  12449--12460.

\bibitem[{Baltru{\v{s}}aitis, Ahuja, and
  Morency(2018)}]{baltruvsaitis2018multimodal}
Baltru{\v{s}}aitis, T.; Ahuja, C.; and Morency, L.-P. 2018.
\newblock Multimodal machine learning: A survey and taxonomy.
\newblock \emph{IEEE transactions on pattern analysis and machine
  intelligence}, 41(2): 423--443.

\bibitem[{Baltru{\v{s}}aitis, Mahmoud, and Robinson(2015)}]{action_unit}
Baltru{\v{s}}aitis, T.; Mahmoud, M.; and Robinson, P. 2015.
\newblock Cross-dataset learning and person-specific normalisation for
  automatic action unit detection.
\newblock In \emph{2015 11th IEEE International Conference and Workshops on
  Automatic Face and Gesture Recognition (FG)}, volume~6, 1--6. IEEE.

\bibitem[{Devlin et~al.(2018)Devlin, Chang, Lee, and Toutanova}]{BERT}
Devlin, J.; Chang, M.-W.; Lee, K.; and Toutanova, K. 2018.
\newblock Bert: Pre-training of deep bidirectional transformers for language
  understanding.
\newblock \emph{arXiv preprint arXiv:1810.04805}.

\bibitem[{Eyben et~al.(2015)Eyben, Scherer, Schuller, Sundberg, Andr{\'e},
  Busso, Devillers, Epps, Laukka, Narayanan et~al.}]{egemaps}
Eyben, F.; Scherer, K.~R.; Schuller, B.~W.; Sundberg, J.; Andr{\'e}, E.; Busso,
  C.; Devillers, L.~Y.; Epps, J.; Laukka, P.; Narayanan, S.~S.; et~al. 2015.
\newblock The Geneva minimalistic acoustic parameter set (GeMAPS) for voice
  research and affective computing.
\newblock \emph{IEEE transactions on affective computing}, 7(2): 190--202.

\bibitem[{Han, Chen, and Poria(2021)}]{MMIM}
Han, W.; Chen, H.; and Poria, S. 2021.
\newblock Improving Multimodal Fusion with Hierarchical Mutual Information
  Maximization for Multimodal Sentiment Analysis.
\newblock In \emph{Proceedings of the 2021 Conference on Empirical Methods in
  Natural Language Processing}, 9180--9192.

\bibitem[{Hazarika, Zimmermann, and Poria(2020)}]{MISA}
Hazarika, D.; Zimmermann, R.; and Poria, S. 2020.
\newblock Misa: Modality-invariant and-specific representations for multimodal
  sentiment analysis.
\newblock In \emph{Proceedings of the 28th ACM International Conference on
  Multimedia}, 1122--1131.

\bibitem[{Li et~al.(2020)Li, Li, Duan, Zheng, and Zhao}]{TPFN}
Li, B.; Li, C.; Duan, F.; Zheng, N.; and Zhao, Q. 2020.
\newblock Tpfn: Applying outer product along time to multimodal sentiment
  analysis fusion on incomplete data.
\newblock In \emph{European Conference on Computer Vision}, 431--447. Springer.

\bibitem[{Liang, Zadeh, and Morency(2022)}]{NewSurvey}
Liang, P.~P.; Zadeh, A.; and Morency, L.-P. 2022.
\newblock Foundations and Recent Trends in Multimodal Machine Learning:
  Principles, Challenges, and Open Questions.
\newblock \emph{arXiv preprint arXiv:2209.03430}.

\bibitem[{Liu et~al.(2022)Liu, Yuan, Mao, Liang, Yang, Qiu, Cheng, Li, Xu, and
  Gao}]{CH-SIMSv2}
Liu, Y.; Yuan, Z.; Mao, H.; Liang, Z.; Yang, W.; Qiu, Y.; Cheng, T.; Li, X.;
  Xu, H.; and Gao, K. 2022.
\newblock Make Acoustic and Visual Cues Matter: CH-SIMS v2. 0 Dataset and
  AV-Mixup Consistent Module.
\newblock In \emph{INTERNATIONAL CONFERENCE ON MULTIMODAL INTERACTION},
  247--258.

\bibitem[{Liu et~al.(2018)Liu, Shen, Lakshminarasimhan, Liang, Zadeh, and
  Morency}]{LMF}
Liu, Z.; Shen, Y.; Lakshminarasimhan, V.~B.; Liang, P.~P.; Zadeh, A.~B.; and
  Morency, L.-P. 2018.
\newblock Efficient Low-rank Multimodal Fusion With Modality-Specific Factors.
\newblock In \emph{Proceedings of the 56th Annual Meeting of the Association
  for Computational Linguistics (Volume 1: Long Papers)}, 2247--2256.

\bibitem[{Mao et~al.(2022)Mao, Yuan, Xu, Yu, Liu, and Gao}]{M-SENA}
Mao, H.; Yuan, Z.; Xu, H.; Yu, W.; Liu, Y.; and Gao, K. 2022.
\newblock M-SENA: An Integrated Platform for Multimodal Sentiment Analysis.
\newblock \emph{arXiv preprint arXiv:2203.12441}.

\bibitem[{Rahman et~al.(2020)Rahman, Hasan, Lee, Zadeh, Mao, Morency, and
  Hoque}]{BERT_MAG}
Rahman, W.; Hasan, M.~K.; Lee, S.; Zadeh, A.~B.; Mao, C.; Morency, L.-P.; and
  Hoque, E. 2020.
\newblock Integrating multimodal information in large pretrained transformers.
\newblock In \emph{Proceedings of the 58th Annual Meeting of the Association
  for Computational Linguistics}, 2359--2369.

\bibitem[{Soleymani et~al.(2017)Soleymani, Garcia, Jou, Schuller, Chang, and
  Pantic}]{soleymani2017survey}
Soleymani, M.; Garcia, D.; Jou, B.; Schuller, B.; Chang, S.-F.; and Pantic, M.
  2017.
\newblock A survey of multimodal sentiment analysis.
\newblock \emph{Image and Vision Computing}, 65: 3--14.

\bibitem[{Thiemann, Ito, and Vincent(2013)}]{DEMAND}
Thiemann, J.; Ito, N.; and Vincent, E. 2013.
\newblock The diverse environments multi-channel acoustic noise database
  (demand): A database of multichannel environmental noise recordings.
\newblock In \emph{Proceedings of Meetings on Acoustics ICA2013}, volume~19,
  035081. Acoustical Society of America.

\bibitem[{Tourapis(2002)}]{EPZS}
Tourapis, A.~M. 2002.
\newblock Enhanced predictive zonal search for single and multiple frame motion
  estimation.
\newblock In \emph{Visual Communications and Image Processing 2002}, volume
  4671, 1069--1079. SPIE.

\bibitem[{Yu et~al.(2020)Yu, Xu, Meng, Zhu, Ma, Wu, Zou, and Yang}]{CH-SIMS}
Yu, W.; Xu, H.; Meng, F.; Zhu, Y.; Ma, Y.; Wu, J.; Zou, J.; and Yang, K. 2020.
\newblock {CH}-{SIMS}: A {C}hinese Multimodal Sentiment Analysis Dataset with
  Fine-grained Annotation of Modality.
\newblock In \emph{Proceedings of the 58th Annual Meeting of the Association
  for Computational Linguistics}, 3718--3727. Online: Association for
  Computational Linguistics.

\bibitem[{Yu et~al.(2021)Yu, Xu, Yuan, and Wu}]{self-mm}
Yu, W.; Xu, H.; Yuan, Z.; and Wu, J. 2021.
\newblock Learning Modality-Specific Representations with Self-Supervised
  Multi-Task Learning for Multimodal Sentiment Analysis.
\newblock In \emph{Proceedings of the AAAI Conference on Artificial
  Intelligence}, volume~35, 10790--10797.

\bibitem[{Yuan et~al.(2021)Yuan, Li, Xu, and Yu}]{TFR-NET}
Yuan, Z.; Li, W.; Xu, H.; and Yu, W. 2021.
\newblock Transformer-based Feature Reconstruction Network for Robust
  Multimodal Sentiment Analysis.
\newblock In \emph{Proceedings of the 29th ACM International Conference on
  Multimedia}, 4400--4407.

\bibitem[{Zadeh et~al.(2017{\natexlab{a}})Zadeh, Chen, Poria, Cambria, and
  Morency}]{TFN}
Zadeh, A.; Chen, M.; Poria, S.; Cambria, E.; and Morency, L.-P.
  2017{\natexlab{a}}.
\newblock Tensor Fusion Network for Multimodal Sentiment Analysis.
\newblock In \emph{Proceedings of the 2017 Conference on Empirical Methods in
  Natural Language Processing}, 1103--1114.

\bibitem[{Zadeh et~al.(2017{\natexlab{b}})Zadeh, Chong~Lim, Baltrusaitis, and
  Morency}]{landmarks}
Zadeh, A.; Chong~Lim, Y.; Baltrusaitis, T.; and Morency, L.-P.
  2017{\natexlab{b}}.
\newblock Convolutional experts constrained local model for 3d facial landmark
  detection.
\newblock In \emph{Proceedings of the IEEE International Conference on Computer
  Vision Workshops}, 2519--2528.

\bibitem[{Zadeh et~al.(2016)Zadeh, Zellers, Pincus, and Morency}]{MOSI}
Zadeh, A.; Zellers, R.; Pincus, E.; and Morency, L.-P. 2016.
\newblock Multimodal sentiment intensity analysis in videos: Facial gestures
  and verbal messages.
\newblock \emph{IEEE Intelligent Systems}, 31(6): 82--88.

\bibitem[{Zadeh et~al.(2018)Zadeh, Liang, Poria, Cambria, and Morency}]{MOSEI}
Zadeh, A.~B.; Liang, P.~P.; Poria, S.; Cambria, E.; and Morency, L.-P. 2018.
\newblock Multimodal language analysis in the wild: Cmu-mosei dataset and
  interpretable dynamic fusion graph.
\newblock In \emph{Proceedings of the 56th Annual Meeting of the Association
  for Computational Linguistics (Volume 1: Long Papers)}, 2236--2246.

\end{thebibliography}
\end{document}